\documentclass{article}
\usepackage[utf8]{inputenc}
\usepackage[english]{babel}
\usepackage[margin=0.8in]{geometry}
\usepackage{cite}
\usepackage{graphicx}
\usepackage{amssymb,amsfonts,dsfont,mathrsfs,amsthm,mathtools}
\newcommand{\diff}[1]{\text{d}#1}
\newcommand{\Diff}[1]{\mathscr{D}#1}

\newcommand*{\diag}{\operatorname{diag}}

\newcommand{\Lag}{\mathscr{L}}

\title{Is there any symmetry left in gravity theories with explicit Lorentz violation?}
\author{Yuri Bonder and Crist\'obal Corral\\ \\
Instituto de Ciencias Nucleares, Universidad Nacional Aut\'onoma de M\'exico\\ Apartado Postal 70-543, Ciudad de M\'exico, 04510, M\'exico\\ \\
bonder@nucleares.unam.mx; cristobal.corral@correo.nucleares.unam.mx}        
\date{}                                   % Activate to display a given date or no date

\begin{document}
\maketitle

\section*{Abstract}
It is well known that a theory with explicit Lorentz violation is not invariant under diffeomorphisms. On the other hand, for geometrical theories of gravity, there are alternative transformations, which can be best defined within the first-order formalism, and that can be regarded as a set of improved diffeomorphisms. These symmetries are known as local translations and, among other features, they are Lorentz covariant off shell. It is thus interesting to study if theories with explicit Lorentz violation are invariant under local translations. In this work, an example of such a theory, known as the minimal gravity sector of the Standard Model Extension, is analyzed. Using a robust algorithm, it is shown that local translations are not a symmetry of the theory. It remains to be seen if local translations are spontaneously broken under spontaneous Lorentz violation, which are regarded as a more natural alternative when spacetime is dynamic.

\section{Introduction}

Conventional theories of gravity, when geometrical, are invariant under diffeomorphisms (Diff) and local Lorentz transformations (LLT). In addition, such theories are invariant under the so-called local translations (LT), which can be regarded as improved Diff in the sense that they are fully Lorentz covariant, among other propeties \cite{Hehl:1994ue}. As expected, for theories that are invariant under LLT, on shell, invariance under Diff implies invariance under LT and vice versa \cite{Hehl:1976kj,Blagojevic:2002du}. Thus, in light of this result, it is interesting to study the logical relation between Diff and LT when LLT is explicitly broken. This is the main goal of this work. Clearly, a symmetry is a very powerful tool that simplifies calculations and provides conceptual clarity, therefore, elucidating if there is a symmetry left in theories with explicit Lorentz violation can be extremely relevant.

General relativity is thought to be a low energy limit of a more fundamental theory of gravity that incorporates the quantum principles \cite{Burgess}. Several interesting quantum gravity approaches have been developed. The leading candidates include string theory, loop quantum gravity, spin networks, noncommutative geometry, causal sets, and causal dynamical triangulations (see the corresponding contributions in Ref.~\cite{Oriti}). Notably, even though this theory is still unknown, research fields like quantum cosmology are making steady progress by appealing to effects inspired by quantum gravity \cite{QC1,QC2}.

On the other hand, there is an approach to quantum gravity where the priority is to make contact with experiments and which is known as quantum gravity phenomenology. Within quantum gravity phenomenology, it is customary to look for traces of Lorentz violation. The motivation to do this stems from the fact that within the most prominent approaches to quantum gravity, it has been argued that LLT may not be a fundamental symmetry \cite{LVStrings,LVStrings1,LVLQG,LVLQG1,LVnoncommutativity}, and thus, it may be possible to look for empirical traces of quantum gravity by searching for violations of LLT. Inspired by this possibility, a parametrization of Lorentz violation based on effective field theory has been developed \cite{SME1,SME2}, which is known as the Standard Model Extension (SME).

The SME action contains Lorentz-violating extensions to all sectors of conventional physics, including general relativity \cite{SME3}; this latter sector is known as the gravity sector. Moreover, within the gravity sector, the part that produces second-order field equations for the metric is called minimal gravity sector, and, using a post-Newtonian expansion \cite{Test1}, it has been tested with several interesting experiments including atom interferometry \cite{Test2}, frame dragging \cite{Test3}, lunar ranging \cite{LLR1,LLR2}, pulsar timing \cite{Pulsar1,Pulsar2}, and planetary motion \cite{Planetary}. There are also tests in the context of cosmology \cite{Test4}. Other experiments related to the SME are reported in Ref.~\cite{DataTables}.

It should be stressed that the dominant position in the SME community is that, in the presence of gravity, Lorentz violation must occur spontaneously. This is assumed to deal with the severe restrictions arising from the contracted Bianchi identity \cite{SME3}, i.e., the fact that the Einstein tensor is divergence free. This, in turn, is closely related to the invariance under Diff. In addition, some relations between Diff and LLT are known in the literature. For example, it has been shown that spontaneous violation of Diff implies spontaneous violation of LLT, and vice versa \cite{Nogo1,Nogo2,Nogo3}. In these regards some surprising results have already been uncovered by using LT. In particular, in the unimodular theory of gravity \cite{Bonder:2018mfz}, explicit breaking of Diff produces a breakdown of LT but, contrary to the expectations, LLT is unaffected \cite{Nosotros}. Thus, it is interesting to analyze the fate of the LT when LLT is spontaneously broken. Here, however, attention is restricted to explicit symmetry breaking in the minimal gravity sector of the SME. This assumption is adopted for simplicity, but also since the dynamics associated with spontaneous Lorentz violation may spoil the Cauchy initial value formulation \cite{Carlitos}. Moreover, this setup allows one to study if torsion, which modifies the Bianchi identities, can relax the restrictions that have driven the SME community to consider spontaneous Lorentz violation. Other alternatives to deal with these restrictions include a St\"uckelberg-like mechanism \cite{Bluhm} and the use of Finsler geometries \cite{Finsler}.

\section{Gauge theories of gravity and local translations}\label{GGT}

Local translations can be best defined in gauge theories of gravity \cite{Hehl:1994ue,Obukhov:2006gea,Blagojevic:2013xpa} that consider two independent gravitational fields: the tetrad $e^{a}{}_\mu$ and the Lorentz connection $\omega^{ab}{}_\mu=-\omega^{ba}{}_\mu$. This setup includes the well-known Poincar\'e Gauge Theories \cite[Chapter 3]{Blagojevic:2002du}. Here, spacetime indices are represented by Greek letters and tangent-space indices with Latin characters; the summation convention on repeated indices is understood. The four-dimensional spacetime metric $g_{\mu\nu} $ is related to the tetrad by $g_{\mu\nu} = \eta_{ab} e^{a}{}_\mu e^{b}{}_\nu$, where $\eta_{ab}=\diag\left(-,+,+,+\right)$. Note that $\eta_{ab}$ and its inverse, $\eta^{ab}$, can be used to lower and raise tangent-space indices, and that the tetrad and its inverse, $E^{\mu}{}_a$, which is such that $e^{a}{}_\mu E^{\mu}{}_b = \delta^a_b$ and $e^{a}{}_\mu E^{\nu}{}_a = \delta^\mu_\nu$, can be used to map spacetime indices to tangent-space indices and vice versa. In addition, the Lorentz connection and the spacetime connection $ \Gamma^{\lambda}{}_{\mu\nu} $ are related through the tetrad postulate
\begin{equation}\label{tetpost}
 \partial_\mu e^{a}{}_\nu + \omega^{a}{}_{b\mu} e^{b}{}_\nu = \Gamma^{\lambda}{}_{\mu\nu} e^{a}{}_\lambda.
 \end{equation}
The left-hand side of this equation can be written as $\Diff{}_\mu e^{a}{}_\nu$ where $\Diff{}_\mu$ is the covariant derivative with respect to the Lorentz connection\footnote{The covariant derivative used here differs from the operator widely used in SME papers (e.g., Ref.~\cite{SME3}), which is represented by $\text{D}_\mu$, in that, when acting on a tensor, $\Diff{}_\mu$ does not add a connection term for each spacetime index. These two operators are discussed in Ref.~\cite{Zanelli:2016cs} in a notation that does not coincide with the one used here.}.

Curvature and torsion can be derived from the tetrad and the Lorentz connection through Cartan's structure equations
\begin{align}
\label{curvaturedef}
 \frac{1}{2}R^{ab}{}_{\mu\nu} &= \partial_{[\mu}\omega^{ab}{}_{\nu]} + \omega^{a}{}_{c[\mu} \omega^{cb}{}_{\nu]}, \\
\label{torsiondef}
 \frac{1}{2}T^{a}{}_{\mu\nu} &=  \Diff{}_{[\mu}e^{a}{}_{\nu]},
\end{align}
where the squared brackets denote antisymmetrization of the $n$ indices enclosed (with a $1/n!$ factor). Clearly, $R^{ab}{}_{\mu\nu}=R^{[ab]}{}_{[\mu\nu]}$ and $T^{a}{}_{\mu\nu}=T^{a}{}_{[\mu\nu]}$ and, from Eqs.~\eqref{tetpost} and~\eqref{torsiondef}, it can be verified that $T^{a}{}_{\mu\nu}/2=\Gamma^{\lambda}{}_{[\mu\nu]} e^{a}{}_\lambda$. Furthermore, the Bianchi identities take the form 
\begin{align}
\Diff{}_{[\mu} R^{ab}{}_{\nu\lambda]} - T^{c}{}_{[\mu\nu} R^{ab}{}_{\lambda]\rho} E^{\rho}{}_c &= 0,\\
\label{Bianchi1}\Diff{}_{[\mu} T^{a}{}_{\nu\lambda]} - T^{b}{}_{[\mu\nu} T^{a}{}_{\lambda]\rho} E^{\rho}{}_b &= R^{a}{}_{[\mu\nu\lambda]}.
\end{align}

As usual, (active) infinitesimal Diff are implemented by the Lie derivative. Since the tetrad and the Lorentz connection are $1$-forms, it reads
\begin{align}
\label{Diff}
 \mbox{Diff} &= \begin{cases}
 \delta_{\rm Diff}(\rho) e^{a}{}_\mu &= \rho^\nu \partial_\nu  e^{a}{}_\mu + \partial_\mu\rho^\nu e^{a}{}_\nu ,\\
 \delta_{\rm Diff}(\rho) \omega^{ab}{}_\mu &= \rho^\nu \partial_\nu\omega^{ab}{}_\mu + \partial_\mu\rho^\nu \omega^{ab}{}_\nu.
 \end{cases}
\end{align}
Moreover, under LLT the tetrad and Lorentz connection respectively transform as a vector and gauge connection, that is
\begin{align}
\label{LLT}
 \mbox{LLT} &= \begin{cases}
 \delta_{\rm LLT}(\lambda) e^{a}{}_\mu &= -\lambda^{a}{}_b e^{b}{}_\mu,\\
 \delta_{\rm LLT}(\lambda) \omega^{ab}{}_\mu &= \Diff{}_\mu\lambda^{ab}.
 \end{cases}
\end{align}
The fact that Diff involves partial derivatives and not covariant derivatives under LLT already suggests that Diff are not Lorentz covariant. However, this observation also points to its cure: define a transformation replacing the partial derivative by a covariant derivative. This is the most conventional way to introduce the LT \cite{Hehl:1994ue} and the result, for the case where the gauge group is LLT, is
\begin{eqnarray}\label{LT}
 \text{LT} &= \begin{cases}
 \delta_{\rm LT}(\rho) e^{a}{}_\mu\;\, = \Diff{}_\mu\rho^a + \rho^\nu T^{a}{}_{\nu\mu}, \\
 \delta_{\rm LT}(\rho) \omega^{ab}{}_\mu = \rho^\nu R^{ab}{}_{\nu\mu},
 \end{cases}
\end{eqnarray}
where $\rho^a = e^{a}{}_\mu \rho^\mu$. Remarkably, it is easy to verify that, acting on $e^{a}{}_\mu$ or $\omega^{ab}{}_\mu$,
\begin{equation}\label{ltdiff}
 \delta_{\rm Diff}(\rho) = \delta_{\rm LT}(\rho) + \delta_{\rm LLT}(\tilde{\lambda}),
\end{equation}
where
\begin{equation}
\tilde{\lambda}^{ab} = \rho^\mu\omega^{ab}{}_\mu.
\end{equation}
From this relation it can be verified that Diff and LT are equivalent in theories that are invariant under LLT. What is more, if a theory is invariant under two of these symmetries, it has to be invariant under the third. Conversely, this suggests that, if a theory breaks a symmetry, as it is the case of the SME, it should break at least one of the remaining symmetries.

Recall that general relativity is invariant under the two symmetry classes: LLT and Diff. The former acts locally on the tangent space and it thus can be regarded as the gauge (or internal) symmetry of the theory, while the latter connects different spacetime points. Note that, from Eq.~\eqref{ltdiff}, it seems that the LT act in both, the tangent space, through LLT, and the manifold, via the Diff. However, the fact that the gravitational potentials transform under LT as Lorentz tensors, which does not occur under Diff, can be used to argue that they only act on the tangent space.

Now, it is possible to construct theories where the gauge group is different from LLT. Perhaps the most popular examples of such theories are those where the gauge group is de Sitter or anti-de Sitter~\cite{MacDowell,Stelle:1979va,Troncoso:1999pk,Zanelli:2016cs}. Fortunately, the LT definition can be generalized to theories invariant under arbitrary gauge transformations (GT). Thus, for the sake of generality, in the remaining of this section the role of the LLT is played by a general GT. To obtain the generalized LT one simply needs to replace the partial derivatives, in Eqs.~\eqref{Diff}, by covariant derivatives associated to the corresponding gauge group. Equivalently, they can be defined \cite{Obukhov:2007se} as the difference of $\delta_{\rm Diff}(\rho) $ and $\delta_{\rm GT}(\tilde{\lambda})$, for the corresponding $\tilde{\lambda}^{ab} $, thus generalizing Eq.~\eqref{ltdiff}. It is important to remark that the action of LT on the geometrical fields depends on the gauge symmetry and, for theories invariant under generic GT, it does not need to coincide with Eqs.~\eqref{LT}.

Interestingly, an algorithm has been developed \cite{Nosotros} that, starting from the action, allows one to verify if the theory is invariant under LT and GT and, if it does, it gives the corresponding transformations of the dynamical fields (see also Ref.~\cite{Montesinos:2017epa}). What is more, in certain cases, the algorithm leads to the corresponding contracted Bianchi identities and the matter conservation laws. Since the algorithm is closely related to N\"other's theorem, as it can be seen from Eqs.~\eqref{algo1} and \eqref{algo2}, it selects the fundamental symmetries of the theory. In fact, the algorithm's output contains the transformation laws of the dynamical fields under GT and LT; whether the theory is invariant under Diff can be derived from such transformations. In this sense the GT and LT are more fundamental than the Diff. The covariant derivative associated with GT is denoted by $\bar{\Diff{}}_\mu$, and the situation in which GT is the Lorentz group, which leads to Eqs.~\eqref{LT}, arises as a particular case. Also, for simplicity, the algorithm is done in a $4$-dimensional spacetime that has no boundaries, and all dynamical fields besides the tetrad and Lorentz connection, which are denoted by $\Psi$, are assumed to be $0$-forms in a nontrivial representation of the Lorentz group. Notice that $\Psi$ may be Dirac spinors. Also observe that the fact that the formalism is based on the action, and not the Hamiltonian, allows one to disregard all issues related with spacetime foliations.

The basic steps of the algorithm are~(i) consider an action principle
\begin{equation}
S[e^{a}{}_\mu,\omega^{ab}{}_\mu,\Psi]= \int\diff{}^4x\, e\, \Lag(e^{a}{}_\mu,\omega^{ab}{}_\mu,\Psi),
\end{equation}
where $\diff{}^4x\, e $ is the covariant $4$-volume element and $\Lag$ is an arbitrary Lagrangian. (ii)~Perform an arbitrary variation of the action with respect to the dynamical fields
\begin{equation}\label{genvar}
 \delta S = \int\diff{}^4x\, e\, \left(\delta e^{a}{}_\mu F^{\mu}{}_a+ \delta\omega^{ab}{}_\mu F^{\mu}{}_{ab} + \delta\Psi F \right),
\end{equation}
which implicitly defines $F^{\mu}{}_a$,  $F^{\mu}{}_{ab}$, and $F$. Note that $F^{\mu}{}_a$ includes the variation of the volume element, namely, $F^{\mu}{}_a=(1/e)\delta(e\Lag)/\delta e^a_\mu$, and that these objects vanish on shell. Then, (iii)~apply a covariant derivative $\bar{\Diff{}}_\mu$ to $F^{\mu}{}_a$, $F^{\mu}{}_{ab}$, and $F$. Step (iv),~verify if the resulting expressions can be written as linear combinations of $F^{\mu}{}_a$, $F^{\mu}{}_{ab}$, and $F$, where the coefficients can be spacetime functions. If possible, it reads
\begin{align}\label{LTBianchi}
 \bar{\Diff{}}_\mu F^{\mu}{}_a &= \alpha_\mu F^{\mu}{}_a+ \beta^{b}{}_\mu F^{\mu}{}_{ab} + \gamma_a F,\\
 \label{LLTBianchi}
 \bar{\Diff{}}_\mu F^{\mu}{}_{ab} &= \pi_{\mu[a} F^{\mu}{}_{b]} + \theta_\mu F^{\mu}{}_{ab} + \mu_{[ab]} F,\\
 \bar{\Diff{}}_\mu F &= 0, 
\end{align}
where the fact that $F$ is a $4$-form is used. Note that the precise form of the coefficients in Eqs.~\eqref{LTBianchi} and~\eqref{LLTBianchi} depends on the theory, and, in particular, on the gauge group. Step (v),~multiply Eqs.~\eqref{LTBianchi} and~\eqref{LLTBianchi}, respectively, with gauge parameters $\rho^a(x)$ and $\xi^{ab}(x)=-\xi^{ba}(x)$. Finally, (vi)~integrate over spacetime using the appropriate $4$-volume element. At this stage one must use that $\bar{\Diff{}}_\mu $ can be replaced by $\partial_\mu$ when acting on a scalar under the gauge group. Also, the Leibniz rule, Gauss's theorem, and the assumption that there are no spacetime boundaries are utilized to take the equations to the form
\begin{align}
0 &= \int\diff{}^4x\,  \partial_\mu\left(e\,F^{\mu}{}_a \rho^a \right)\nonumber\\
& = \int\diff{}^4x\, e\, [\underbrace{\left(\rho^a \partial_\mu \ln e+\bar{\Diff{}}_\mu\rho^a + \alpha_\mu \rho^a \right)}_{\delta_{\rm LT}(\rho)e^{a}{}_\mu}F^{\mu}{}_a + \underbrace{\rho^{[a}\beta^{b]}{}_\mu}_{\delta_{\rm LT}(\rho)\omega^{ab}{}_\mu} F^{\mu}{}_{ab} + \underbrace{\gamma_a \rho^a}_{\delta_{\rm LT}(\rho)\Psi} F],\label{algo1}\\
 0&= \int\diff{}^4x\, \partial_\mu\left(e\, F^{\mu}{}_{ab} \xi^{ab} \right)\nonumber\\
 & =\int\diff{}^4x\, e\, [\underbrace{-\pi_{\mu b}\xi^{ab}}_{\delta_{\rm GT}(\xi) e^{a}{}_\mu} F^{\mu}{}_a + \underbrace{\left(\xi^{ab} \partial_\mu \ln e+\bar{\Diff{}}_\mu \xi^{ab} + \theta_\mu \xi^{ab}\right)}_{\delta_{\rm GT}(\xi)\omega^{ab}{}_\mu} F^{\mu}{}_{ab} + \underbrace{\mu_{ab}\xi^{ab}}_{\delta_{\rm GT}(\xi)\Psi}F].\label{algo2}
\end{align}
Then, comparing with the action variation~\eqref{genvar}, it is possible to read off the field transformations under LT and GT. On the other hand, if it is impossible to write the covariant derivatives of $F^{\mu}{}_a$ and $F^{\mu}{}_{ab}$ as in Eqs.~\eqref{LTBianchi} and~\eqref{LLTBianchi}, then one can identify which symmetries are broken and which terms are responsible for such breakdowns. In the next section, this algorithm is applied to the minimal gravity sector of the SME.

\section{Explicit Lorentz violation in the gravity sector}

A theory with explicit Lorentz violation is one in which identical experiments done in different inertial frames can produce different results. It is thus easy to imagine \textit{gedanken} and realistic experiments designed to look for such violations. In fact, the Earth's rotation (translation) gives a natural family of instantaneous inertial frames in which Lorentz violation can manifest themselves as signals with a daily (yearly) period. Moreover, there are concrete models that incorporate Lorentz violation to account for astrophysical puzzles like the presence of cosmic rays above the GZK cutoff \cite{PhysRevD.59.116008}. Here attention is restricted to a sector of the SME, which should be regarded as a generic parametrization of Lorentz violation.

The action of the minimal gravity sector of the SME, in the first-order formalism, takes the form $S=S_g+S_m$ where the Lorentz-violating gravitational part is
\begin{equation}\label{lvaction}
 S_g[e^{a}{}_\mu,\omega^{ab}{}_\mu] = \frac{1}{2\kappa}\int\diff{}^4x\,e\,E^{\mu}{}_a E^{\nu}{}_b\left(R^{ab}{}_{\mu\nu}+k_{cd}{}^{ab} R^{cd}{}_{\mu\nu}\right).
\end{equation}
Here $\kappa=8\pi G_N$ is the gravitational coupling constant, and $e=\det e^{a}{}_\mu$. The first term of this action is the Einstein--Hilbert action, and $k_{cd}{}^{ab}$ is a nondynamical $0$-form parametrizing possible Lorentz violations, and whose components are known as the SME coefficients. The SME coefficients are typically expressed in terms of spacetime indices, this can be achieved by using the tetrad and its inverse to translate the Latin indices in $k_{cd}{}^{ab}$ to spacetime indices. From the index symmetries of $R^{ab}{}_{\mu\nu}$ is it clear that $k_{cd}{}^{ab}= k_{[cd]}{}^{[ab]}$. However, from Eq.~\eqref{Bianchi1} it can be seen that, in the presence of torsion, it is not necessary to assume that $k_{a[bcd]}=0$. Thus, there are $36$ independent components of $k_{cd}{}^{ab}$; this should be compared with the $20$ independent components that are present when torsion vanishes. Note that the SME coefficients sensitive to torsion are studied in Ref.~\cite{SME3}. For simplicity, even though it has been shown that there is a York--Gibbons--Hawking term for the minimal gravity SME sector \cite{tpuzzle}, spacetime boundaries are not considered. Also, the SME coefficients are assumed to be extremely small in any relevant reference frame, thus, they should not damage the Cauchy initial value formulation of general relativity \cite{Ringstrom}.

The matter action is
\begin{equation}\label{mattaction}
S_m=\int\diff{}^4x\,e \Lag_m[e^{a}{}_\mu,\omega^{ab}{}_\mu,\Psi],
\end{equation}
where $\Psi$ denotes the matter fields, which are taken to be $0$-forms in a nontrivial representation of the Lorentz group (recall that $\Psi$ includes spinors). As it is customary, the energy-momentum and spin densities are respectively defined by
\begin{align}
{\tau^\mu}_a &=\frac{1}{e}\frac{\delta (e \Lag_m)}{\delta {e^a}_\mu},\\
{\sigma^\mu}_{a b}&=2\frac{\delta \Lag_m}{\delta {\omega^{ab}}_\mu}.
\end{align}
Also, since the gravity action does not depend on $\Psi$, it is possible to define, using the notation of the previous section, $F=\delta \Lag_m/\delta \Psi$. The main assumption on $S_m$ is that it is invariant under Diff and LLT. This leads to two off-shell conservation laws
\begin{align}
(\Diff{}_\mu+T_\mu){\tau^\mu}_a &={T^b}_{\mu\nu}{E^\mu}_a {\tau^\nu}_b+\frac{1}{2}{R^{bc}}_{\mu\nu}{E^\mu}_a{\sigma^\nu}_{bc}-F {E^\mu}_a \partial_\mu \Psi\nonumber\\
&\quad-\frac{1}{2}{E^\mu}_a\omega^{bc}{}_\mu[(\Diff{}_\nu+T_\nu){\sigma^\nu}_{bc}-2g_{\nu\rho}{E^\nu}_{b} {\tau^\rho}_{c}],\label{DiffInv}\\
(\Diff{}_\mu+T_\mu){\sigma^\mu}_{a b}&=2g_{\mu\nu}{E^\mu}_{[a} {\tau^\nu}_{b]}+FJ_{ab}\Psi,\label{LLTInv}
\end{align}
where $T_\mu = T^{a}{}_{\mu\nu} E^{\nu}{}_a$ and $J_{ab}$ are the generators of LLT associated with $\Psi$, i.e., $\delta_{\rm LLT}(\lambda) \Psi = -\lambda^{ab}J_{ab}\Psi/2$. Notice that the presence of a $T_\mu$ term next to every $\Diff{}_\mu$ is closely related to the presence of $\partial_\mu \ln e$ in the transformations that are read off from Eqs.~\eqref{algo1} and \eqref{algo2}.

Equation \eqref{DiffInv} is the generalization of the energy-momentum conservation law. In this equation, the issues associated with Diff invariance, which are mentioned in Sec.~\ref{GGT}, become evident: there is a partial derivative of $\Psi$ and the Lorentz connection appears explicitly, and these two terms are not covariant under LLT. Of course, this is not an issue when Eq.~\eqref{LLTInv} is valid because, when these two equations are put together, the term with the Lorentz connection gets replaced by a term with $J_{ab}$ that combines with the partial derivative to transform covariantly. In addition, since these terms are multiplied by $F$, they vanish on shell. Of course, the LT are precisely built in such a way that the problematic terms do not even arise.

An arbitrary variation of the total action is given by Eq.~\eqref{genvar} with
\begin{align}
 F^{\mu}{}_a &=-{\tau^\mu}_a -\frac{1}{\kappa}\left[{G^{\mu}}_a-\frac{1}{2}{k_{bc}}^{de}{R^{bc}}_{\rho\sigma}\left(2{E^\mu}_e{E^\rho}_a{E^\sigma}_d+{E^\mu}_a{E^\rho}_d{E^\sigma}_e\right)\right],\label{eom LV 1}\\
\label{eom LV 2}F^{\mu}{}_{ab} &= -\frac{1}{2}{\sigma^\mu}_{a b}+\frac{1}{2\kappa}\left[ \left({T^{c}}_{\rho\sigma}+2 {e^c}_\rho T_\sigma\right){E^\rho}_{[a}{E^\sigma}_{b]}{E^\mu}_c +{k_{ab}}^{cd}{T^e}_{\rho\sigma}{E^\rho}_{c}{E^\sigma}_{d}{E^\mu}_e \right. \nonumber\\
&\quad \left.- 2{E^\nu}_{c}{E^\mu}_{d} (\Diff{}_\nu+T_\nu){k_{ab}}^{cd}\right],
\end{align}
where ${G^{a}}_\mu = {R^{bc}}_{\rho\sigma}(\delta^a_b \delta^\rho_\mu  {E^\sigma}_c - {e^{a}}_\mu {E^\rho}_b {E^\sigma}_c/2)$ is the Einstein tensor in the presence of torsion. Thus, on shell, Eq.~\eqref{eom LV 1} is the generalization of the Einstein equation, which is not necessarily symmetric, and Eq.~\eqref{eom LV 2} plays the role of the so-called Cartan equation.

The key step in the algorithm presented above is to take the covariant derivative of Eqs.~\eqref{eom LV 1} and \eqref{eom LV 2}. The result are the contracted Bianchi identities, which, using Eqs.~\eqref{DiffInv} and \eqref{LLTInv}, can be casted into the form
\begin{align}
( \Diff{}_\mu+T_\mu)F^{\mu}{}_a &={T^b}_{\mu\nu}{E^\mu}_a {F^\nu}_b+{R^{bc}}_{\mu\nu}{E^\mu}_a{F^\nu}_{bc}-{E^\mu}_a\left(\partial_\mu\Psi+\frac{1}{2}\omega^{ab}{}_\mu J_{ab}\Psi\right)F\nonumber\\
&\quad-\frac{1}{2\kappa} {R^{bc}}_{\rho\sigma}{E^\mu}_a{E^\rho}_d{E^\sigma}_e \Diff{}_\mu {k_{bc}}^{de},\label{contracted Bianchi 1}\\
( \Diff{}_\mu+T_\mu) F^{\mu}{}_{ab} &= 2g_{\mu\nu}{E^\mu}_{[a} {F^\nu}_{b]}+\frac{1}{\kappa}\left({R^{cd}}_{\mu\nu}{k_{cd[a}}^e {E^\mu}_{b]} {E^\nu}_{e}-{R^{c}}_{[a|\mu\nu|}{k_{b]c}}^{de} {E^\mu}_{d} {E^\nu}_{e}\right).\label{contracted Bianchi 2}
\end{align}
Clearly, the term with $\Diff{}_\mu {k_{ab}}^{cd}$ in Eq.~\eqref{contracted Bianchi 1} breaks LT invariance since it cannot be written as a linear combination of $F^{\mu}{}_a$, $F^{\mu}{}_{ab}$ or $F$. Analogously, in Eq.~\eqref{contracted Bianchi 2}, the terms in the parenthesis break LLT. This answers the central question of this paper: the theory is generically not invariant under any of the symmetries considered. Still, by making the last term in Eq.~\eqref{contracted Bianchi 1} equal to zero, while letting the parenthesis in Eq.~\eqref{contracted Bianchi 2} be nonzero, one can break LLT while the theory is invariant under LT. Clearly, it is also possible to have invariance under LLT while LT is broken. However, tuning the values of the SME coefficients goes against the SME philosophy of keeping them arbitrary until they are constrained by experiments.

Remarkably, from Eq.~\eqref{eom LV 2} it can be realized that ${k_{ab}}^{cd}$ acts as a torsion source, as it occurs in theories with nonminimal scalar couplings (see Ref.~\cite{Fernando} and references therein). Thus, even in vacuum, the presence of ${k_{ab}}^{cd}$ generates torsion. This is very unusual since, in most theories, vacuum torsion vanishes. Moreover, this torsion field could be probed with spinors. In fact, for a Dirac spinor $\psi$, the equation of motion for $\psi$ takes the form \cite{KTR}
\begin{equation}
i{E^\mu}_a \gamma^a\left(\partial_\mu \psi +\frac{i}{4}\tilde{\omega}^{bc}{}_\mu \sigma_{bc} \psi \right)+{E^\mu}_a  A_\mu\gamma_5\gamma^a \psi=0,
\end{equation}
where $\gamma^a$ are the Dirac matrices satisfying Clifford's algebra $\left\{\gamma_a,\gamma_b \right\}=- 2\eta_{ab}$, $\sigma_{ab} = i\gamma_{[a}\gamma_{b]}$, and $\gamma_5=i\gamma^0\gamma^1\gamma^2\gamma^3$. In addition, $\tilde{\omega}^{ab}{}_\mu $ is the torsion-free spin connection (i.e., it satisfies the torsion-free tetrad postulate) and $A^\mu=\epsilon^{\rho\sigma\nu\mu}T_{\rho\sigma\nu}/6$, with $\epsilon_{\mu\nu\rho\sigma}$ the Levi-Civita tensor. Notably, bounds on vacuum torsion like those discussed in Ref.~\cite{KTR}, could be motivated by the presence of torsion in the minimal gravity sector of the SME. On the other hand, if torsion is assumed to be nonzero, one can use the very stringent bounds on the matter sector of the SME (see Ref.~\cite{DataTables}) to put limits on ${k_{ab}}^{cd}$; this is similar to what can be done using field redefinitions in the matter-gravity SME sector \cite{yuri2013}. Furthermore, the conventional SME phenomenology is recovered when torsion and spin density are both set to zero. Notice that the former can be rigorously turned off with a Lagrange multiplier~\cite{Simon}. However, even in this case, it is expected that the LT will not be a symmetry of the theory since the structure of Eqs. \eqref{contracted Bianchi 1} and \eqref{contracted Bianchi 2} does not change by the presence of this Lagrange multiplier.

Finally, as it is mentioned in the introduction, in the SME community, Eq.~\eqref{contracted Bianchi 1} is seen as a strong restriction linking matter, geometry, and $k_{abcd}$. However, when torsion and spin density are considered, this condition is relaxed in the sense that it involves more degrees of freedom of both, the matter and the geometry. This may be a valuable alternative in addition to spontaneous Lorentz violation.

\section{Conclusions\label{sec:discussion}}

In this work, the so-called local translations are studied in a gravity theory with explicit Lorentz violation, which is introduced by a nondynamical $0$-form ${k_{ab}}^{cd}$. It was known that the theory is not diffeomorphism invariant, and, in this work, it is shown that the theory is also not invariant under translational invariance. This is interesting since the local translations can be regarded as improved diffeomorphisms in the sense that they are fully covariant under local Lorentz transformations, thus having the potential to be unaffected by Lorentz violation. Another interesting aspect that becomes evident is that the minimal gravity sector of the SME, in the presence of torsion, has additional coefficients which generate vacuum torsion.

There are interesting issues that could be addressed in future contributions. For example, what happens with the local translations in theories with spontaneous Lorentz violation? In this case ${k_{ab}}^{cd}$ would be dynamical and it thus transforms under all symmetries. Furthermore, even if the local translations are indeed spontaneously broken, there could still be advantages of using them when there is spontaneous Lorentz violation. The ultimate role of these local translations needs to be carefully analyzed but, if some traces of them remain, they could play very important roles.

\section*{Acknowledgements}
The authors acknowledge getting valuable input from D.~Gonz\'alez, A.~Kosteleck\'y, and M.~Montesinos. This research was funded by UNAM-DGAPA-PAPIIT Grant No.~RA101818 and UNAM-DGAPA postdoctoral fellowship.

\pagebreak

\bibliographystyle{plain}
\bibliography{References}

\end{document}